\documentclass[epj]{svjour}
\usepackage{graphics}
\begin{document}

\newcommand{\rt}{\rightarrow}
\newcommand{\etal}{\it et al.\rm}

\newcommand{\ppbar}{p \overline{p}}
\newcommand{\jpsi}{J/\psi}

\newcommand{\ee}{e^+ e^-}


\newcommand{\gpppr}{\gamma\pi^+\pi^-p\overline{p}}
\newcommand{\pppr}{\pi^+\pi^-p\overline{p}}
\newcommand{\chicJ}{\chi_{cJ}}
\newcommand{\gaa}{\gamma\Lambda\overline{\Lambda}}
\newcommand{\chicz}{\chi_{c0}}
\newcommand{\ppb}{p\overline{p}}
\newcommand{\chicJto}{\chi_{cJ} \rightarrow }
\newcommand{\aab}{\Lambda\overline{\Lambda}}
\newcommand{\chico}{\chi_{c1}}
\newcommand{\psp}{\psi(2S)}
\newcommand{\psito}{J/\psi \rightarrow }
\newcommand{\chictto}{\chi_{c2} \rightarrow }
\newcommand{\pspto}{\psi(2S) \rightarrow }
\newcommand{\ssb}{\Sigma^0\overline{\Sigma^0}}
\newcommand{\ra}{\rightarrow}
\newcommand{\psipp}{\pi^+\pi^- J/\psi}
\newcommand{\chict}{\chi_{c2}}
\newcommand{\eff}{\varepsilon}
\newcommand{\BR}{{\cal B}}
\newcommand{\pim}{\pi^-}
\newcommand{\chiczto}{\chi_{c0} \rightarrow }

\title{Recent charmonium results from BES\thanks{Talk presented by
    Jiangchuan Chen}}
\author {Jiangchuan 
Chen\inst{1}  \and Frederick A. Harris\inst{2}
} 
\institute {Institute of High Energy 
Physics, Beijing, China  \and Dept. of Physics and Astronomy,
    University of Hawaii, Honolulu, HI 96822, USA }
\date{\today}

\abstract{ Using 58 million $J/\psi$ decays, we have investigated the
  $p \bar{p}$ invariant mass spectrum in the radiative decay $J/\psi
  \rt \gamma p \bar{p}$ and observe a prominent structure with mass
  near $2m_p$. Fitting with an $S$-wave Breit-Wigner, we obtain a peak
  mass of $M= 1859^{+3}_{-10}(\rm stat)^{+5}_{-25}(\rm sys)~{\rm
  MeV}/c^2$.  $J/\psi \rt \gamma \eta_c$ decays from the same sample
  are used to determine the mass, width, and hadronic branching ratios
  of the $\eta_c$.  From a sample of 14 million $\psp$ events, the
  first observation of $\chicJ$ (J=0,1,2) decays to $\aab$ is made,
  and branching ratios are determined, which are larger than expected
  from the Color Octet Model.  Branching ratios of $K_s^0K_L^0$ in
  both $\psi(2S)$ and $J/\psi$ decays are measured, and a more
  than four sigma deviation from the pQCD predicted "12\% rule" is
  observed.  In $\psi(3770)$ decays, evidence for the non-$D\bar{D}$
  decay to $\pi^+\pi^-J/\psi$ is observed.  } \PACS{ {13.20.Gd}{Decays
  of J/psi, Upsilon, and other quarkonia} }
\maketitle
\section{Introduction}

The Beijing Spectrometer (BES) is a general purpose
sole-
noidal
detector at the Beijing Electron Positron Collider (BEPC).  BEPC
operates in the center of mass (cm) energy range from 2 to 5 GeV with a
luminosity at the $J/\psi$ energy of approximately $ 5 \times 10^{30}$
cm$^{-2}$s$^{-1}$.  BES (BESI) is described in detail in Ref.
\cite{bes1}, and the upgraded BES detector (BESII) is described in
Ref. \cite{bes2}.  This paper presents some recent results; details
may be found in the references.

\section{\boldmath Studies of $J/\psi \rt \gamma p \bar{p}$}

There is evidence for anomalous behavior in the proton-antiproton
system very near the $M_{\ppbar}=2m_p$ mass threshold.  The cross
section for $\ee\rt {\rm hadrons}$ has a narrow dip-like structure at a
cm energy of $\sqrt{s} \simeq 2m_p c^2$
\cite{FENICE}.  In addition, the proton's time-like magnetic
form-factor, determined from high statistics measurements of the
$\ppbar\rt\ee$ annihilation process, exhibits a very steep fall-off
just above the $\ppbar$ mass threshold \cite{LEAR}.  These data are
suggestive of a narrow, $S$-wave triplet $\ppbar$ resonance
with $J^{PC} = 1^{--}$ and mass near $M_{\ppbar}\simeq 2m_p$.  

\begin{figure}[htb]
\resizebox{0.41\textwidth}{!}{\includegraphics*{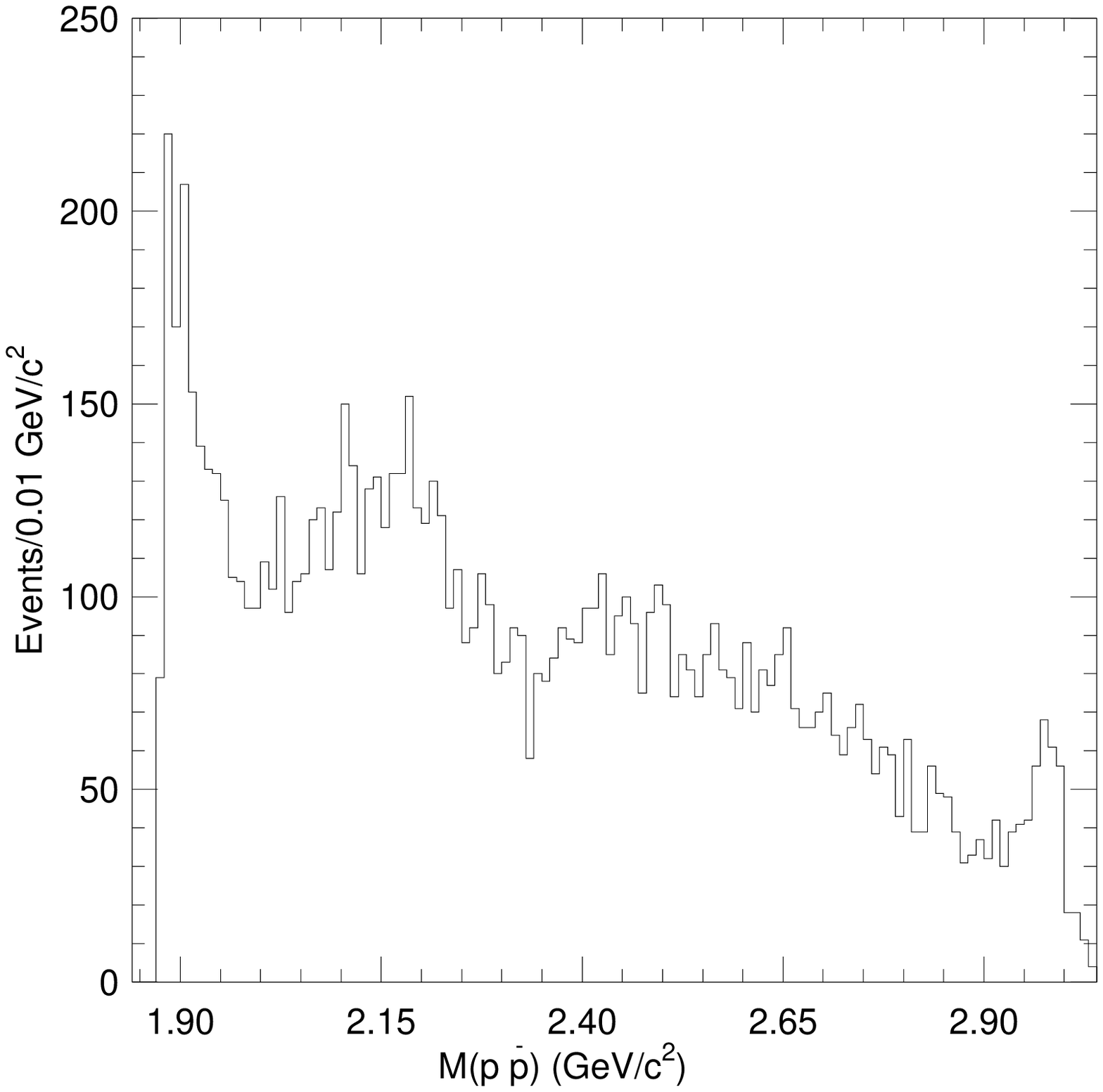}}
\caption{
The $\ppbar$ invariant mass distribution for the
$\jpsi\rt\gamma \ppbar$ event sample}
\label{fig:gpp}
\end{figure}

\begin{figure}[htb]
\resizebox{0.41\textwidth}{!}{\includegraphics*{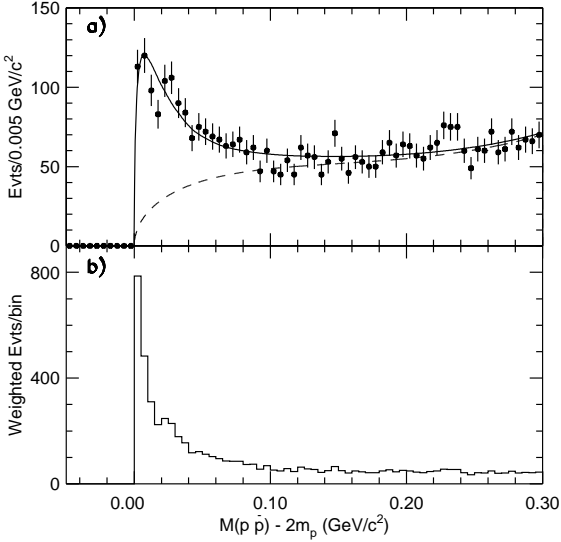}}
\caption{
(a) The near-threshold $M_{\ppbar}-2m_p$ distribution for
the $\gamma\ppbar$ event sample. The solid curve   
is the result of the fit;
the dashed curve shows the fitted background function.
(b) The  $M_{\ppbar}-2m_p$ distribution with events weighted
by $q_0/q$.
}
\label{fig:2pg_thresh_fit}
\end{figure}

Using 58 million $\jpsi$ decays, we have investigated the invariant
mass spectrum of $\ppbar$ pairs in the radiative process
$\jpsi\rt\gamma\ppbar$ and observe a peak corresponding to $J/\psi \rt
\gamma \eta_c$ at high mass and a prominent structure with mass near $2m_p$,
as shown in Fig.~\ref{fig:gpp}.  
Figure~\ref{fig:2pg_thresh_fit}(a) shows the threshold region for the
 selected $J/\psi \rt \gamma \ppbar$ events.
The solid curve is the result of a fit using an acceptance-weighted
 $S$-wave Breit-Wigner function to represent the low-mass enhancement
 plus the dashed curve to represent the background, primarily 
due to $\jpsi\rt\pi^0\ppbar$ where one of
the photons from the $\pi^0\rt\gamma\gamma$ is missed. The fit yields
a peak mass of $M= 1859^{+3}_{-10}(\rm
stat)^{+5}_{-25}(\rm sys)~{\rm MeV}/c^2$ and a full width of $\Gamma <
 30$~MeV. Here the systematic errors include errors determined  by generating
 Monte Carlo samples with below threshold peak masses and measuring the
 shift in the
 output fit masses.  


Further evidence that the peak mass is below the $2m_p$ threshold is
provided by Fig.~\ref{fig:2pg_thresh_fit}(b), which shows the
$M_{\ppbar}-2m_p$ distribution when the threshold behavior is removed
by weighting each event by $q_o/q$, where q is the proton momentum in
the $\ppbar$ restframe and $q_o$ is the value for $M_{\ppbar} = 2$
GeV/$c^2$. The sharp and monotonic increase at threshold can only occur
for an $S$-wave BW function when the peak mass is below 2$m_p$.
A $P$-wave BW yields a fit which is worse than the $S$-wave fit, but
still acceptable.  The structure is
not consistent with the properties of any known particle.
More detail may be found in Ref.~\cite{ppbar}.

\section{\boldmath   $\eta_c$ Parameters}

The mass and width of the $\eta_c$ are rather poorly known; the
confidence level for the Particle Data Group (PDG) weighted average mass is only 0.001
\cite{PDG}. Previously BES measured the $\eta_c$ mass using the BESI
4.02 million $\psi(2S)$ sample and obtained $M_{\eta_c} = (2975.8 \pm 3.9
\pm 1.2)~{\rm MeV}$ \cite{etac1}. BES also used 7.8 million BESI $J/\psi$
events and obtained $M_{\eta_c} = (2976.6 \pm 2.9 \pm 1.3)$ MeV
\cite{etac2}. \ For the two data sets combined, $M_{\eta_c} = (2976.3
\pm 2.3 \pm 1.2)$ MeV and the total width $\Gamma_{\eta_c} = (11.0 \pm
8.1 \pm 4.1)$ MeV {\cite{etac2}.

Here, the mass and width have been determined using our BESII 58 million $J/\psi$
event sample.  We use the channels $J/\psi \rt \gamma \eta_c$, with $\eta_c
\rt p \bar{p}$, $K^+ K^- \pi^+ \pi^-$, $\pi^+ \pi^- \pi^+ \pi^-$,
$K^{\pm} K^o_S \pi^{\mp}$, and $\phi \phi$.  Events are selected using
particle identification and kinematic fitting. Figs.~\ref{fig:eta1} and
\ref{fig:eta2} show the mass distributions in the $\eta_c$ mass region for
$J/\psi \rt \gamma \eta_c$, $\eta_c \rt p \bar{p}$ and $\eta_c \rt
K^+ K^- \pi^+ \pi^-$, respectively.  Fitting the five decay
channels simultaneously, we obtain  $M_{\eta_c} = (2977.5
\pm 1.0 \pm 1.2)$ MeV and $\Gamma_{\eta_c} = (17.0 \pm 3.7 \pm 7.4)$
MeV.  The results for the mass and width are compared with
previous measurements, including previous BES measurements, in
Figs.~\ref{fig:mass} and \ref{fig:width}.  The results are in good
agreement with previous BES measurements and the PDG fit values.
More detail on this analysis may be found in Ref. \cite{etanew}.

\begin{figure}[htb]
  \resizebox{0.41\textwidth}{!}{\includegraphics*{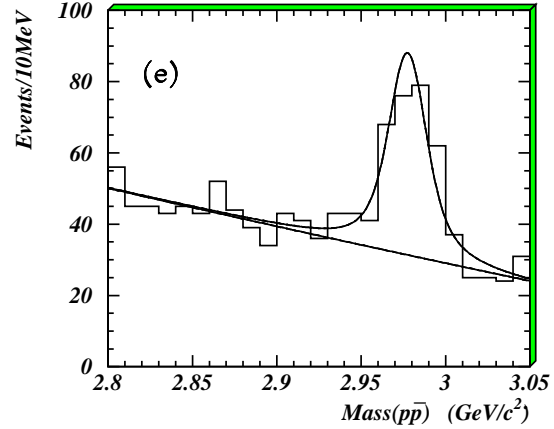}}
     \caption{The  $m_{p\bar{p}}$ invariant mass distribution in the
       $\eta_c$ region.}
     \label{fig:eta1}
\end{figure}

\begin{figure}[htb]
  \resizebox{0.41\textwidth}{!}{\includegraphics*{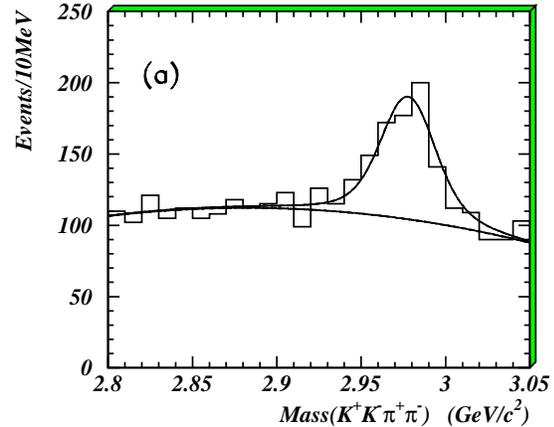}}
     \caption{The  $m_{K^+K^-\pi^+\pi^-}$ invariant mass distribution
       in the $\eta_c$ region.}
     \label{fig:eta2}
\end{figure}

\begin{figure}[htb]
  \resizebox{0.41\textwidth}{!}{\includegraphics*{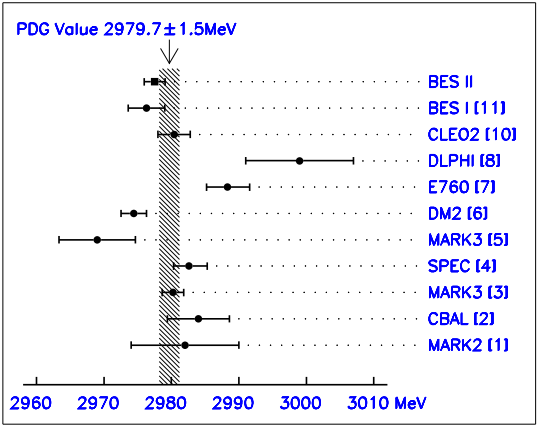}}
     \caption{Mass measurements of the $\eta_c$ meson.}
     \label{fig:mass}
\end{figure}

\begin{figure}[htb]
  \resizebox{0.41\textwidth}{!}{\includegraphics*{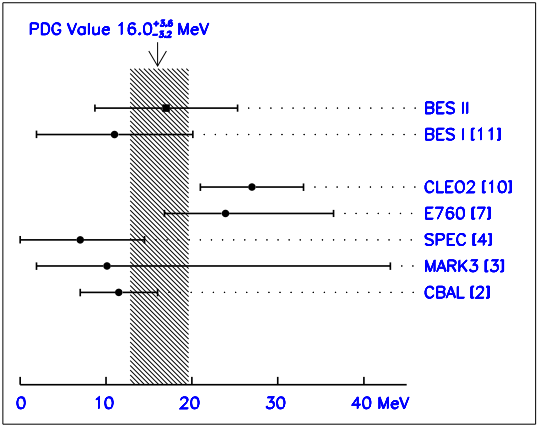}}
     \caption{Width measurements of the $\eta_c$ meson.}
     \label{fig:width}
\end{figure}

The numbers of $\eta_c$ events determined
from the fit and the corresponding product branching ratios, by decay channel,
are listed in Table~\ref{Fit-events}.
Using the branching fraction $Br(J/\psi\to\gamma\eta_c)=(1.3\pm0.4)\%$~\cite{PDG}, 
preliminary $\eta_c$ branching fractions can be obtained. Table~\ref{etac-frac}
shows the BES results together with the PDG~\cite{PDG} values.
The BES $Br(\eta_c\to \phi\phi)$ is smaller than
the current PDG value of $(7.1\pm 2.8) \times 10^{-3}$ and is consistent
with Belle~\cite{belle2} and DM2~\cite{dm2} measurements within errors.
Details may be found in Ref.~\cite{etacbr}.

\begin{table}[htb]
\caption{Number of $\eta_c$ events and corresponding branching ratios
  for the individual channels. Preliminary.} \label{Fit-events}
\begin{center}
\begin{tabular}{|l|c|c|}
\hline
Process & Events & Product of  \\
$J/\psi\to\gamma\eta_c,$    & detected   &  branching ratios                     \\ \hline
$\eta_c \to K^+K^-\pi^+\pi^-$      & $ 413\pm 54$ &
$(1.5\pm0.2\pm 0.2)\times10^{-4}$ \\ \hline
$\eta_c \to \pi^+\pi^-\pi^+\pi^-$  & $ 542\pm 75$ & $(1.3\pm0.2\pm0.4)\times10^{-4}$ \\ \hline
$\eta_c \to K^\pm K_{S}^{0}\pi^\mp$& $ 609\pm 71$ & $(2.2\pm0.3\pm0.5)\times10^{-4}$ \\ \hline
$\eta_c \to \phi\phi$              & $ 357\pm 64$ & $(3.3\pm0.6\pm0.6)\times10^{-5}$ \\ \hline
$\eta_c \to p\bar{p}$              & $ 213\pm 33$ & $(1.9\pm0.3\pm0.3)\times10^{-5}$ \\ \hline
\end{tabular}
\end{center}
\end{table}

\begin{table}[htbp]
\caption{Branching fractions of the $\eta_c$. Preliminary.}
\label{etac-frac}
\begin{center}
\begin{tabular}{|l|c|c|}
\hline
Process                                & BES(\%)        & PDG02(\%) \cite{PDG}          \\ \hline
$Br(\eta_c\to K^+K^-\pi^+\pi^-)$       & $1.2\pm 0.4$   & $2.0^{+0.7}_{-0.6}$         \\ \hline
$Br(\eta_c\to \pi^+\pi^-\pi^+\pi^-)$   & $1.0\pm 0.5$   & $1.2\pm0.4$   \\ \hline
$Br(\eta_c\to K^\pm K_{S}^{0}\pi^\mp)$ & $1.7\pm 0.7$   & $\frac{1}{3}(5.5\pm1.7)$ \\ \hline
$Br(\eta_c\to \phi\phi)$               & $0.25\pm 0.09$ & $0.71\pm0.28$   \\ \hline
$Br(\eta_c\to p\bar{p})$               & $0.15\pm 0.06$ & $0.12\pm0.04$   \\ \hline
\hline
\end{tabular}
\end{center}
\end{table}

\section{\boldmath $\chi_J \rt \Lambda \overline{\Lambda}$}

The lowest Fock state expansion (color singlet mechanism, CSM) of
charmonium states is insufficient to describe P-wave quarkonium
decays. Instead, the next higher Fock state (color octet mechanism,
COM) plays an important role~\cite{so,width}.  A calculation of the
partial width of $\chicJto \ppb$ using the COM and a carefully
constructed nucleon wave function~\cite{wong}, obtains results in
reasonable agreement with measurements~\cite{PDG}.  Generalizing the
nucleon wave function to other baryons, the partial widths of many
other baryon anti-baryon pairs can be predicted. Among these predictions, the
partial width of $\chicJto \aab$ is about half of that of $\chicJto
\ppb$~(J=1,2)~\cite{wong}.



Using 14
million $\psi(2S)$ events and making
a scatter plot of the $\pi^+\overline{p}$
versus the $\pi^-p$ invariant mass for events with $\pppr$ mass
between $3.38$~GeV/$c^2$ and $3.60$~GeV/$c^2$, a clear $\aab$ signal
is observed.
After requiring that both the $\pi^+\overline{p}$ and the $\pi^-p$
mass lie within twice the mass resolution around the nominal $\Lambda$
mass, the $\aab$ invariant mass distribution shown in
Fig.~\ref{maafit} is obtained. There are clear $\chicz$, $\chico$, and
$\chictto \aab$ signals. The highest peak around the $\psp$ mass is
due to $\pspto \aab$ with a fake photon.


\begin{figure}[htb]
  \resizebox{0.41\textwidth}{!}{\includegraphics*{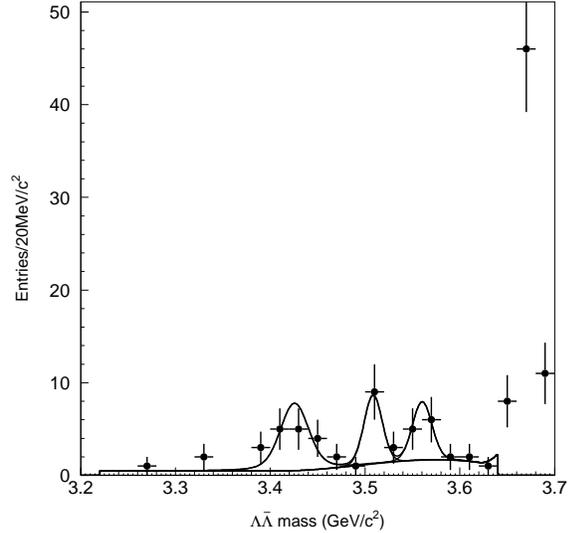}}
\caption{Mass distribution of $\gaa$ candidates fitted with three
resolution smeared Breit-Wigner functions and background, as
described in the text.}
\label{maafit}
\end{figure}

Background from non $\aab$ events is estimated from the $\Lambda$ mass
sidebands. The background from channels with
$\aab$ production, including $\pspto \aab$, $\pspto \ssb$, $\pspto
\Lambda \overline{\Sigma^0} + c.c.$, etc. are
simulated by Monte Carlo.

Fixing the $\chicz$, $\chico$ and $\chict$ mass resolutions at their
Monte Carlo predicted values, and fixing the widths of the three
$\chicJ$ states to their world average values~\cite{PDG},  the mass
spectrum (Fig.~\ref{maafit}) was fit  between 3.22 and
3.64~$\hbox{GeV}/c^2$ with three Breit-Wigner functions
folded with Gaussian resolutions and background, including a linear
term representing the non $\aab$ background and a component
representing the $\aab$ background.
Fig.~\ref{maafit} shows the fit
result.
The branching ratios of $\chicJto \aab$ obtained are
\[ \BR(\chicz \rightarrow \aab)
     = (4.7^{+1.3}_{-1.2} \pm 1.0)\times 10^{-4} ,\]
\[ \BR(\chico \rightarrow \aab)
     = (2.6^{+1.0}_{-0.9} \pm 0.6)\times 10^{-4} ,\]
\[ \BR(\chict \rightarrow \aab)
     = (3.3^{+1.5}_{-1.3} \pm 0.7)\times 10^{-4} ,\]
where the first errors are statistical and the second are 
systematic. 

The results are 
in contradiction with the expectations from Ref.~\cite{wong},
although the errors are large.
There is no prediction for
$\BR(\chiczto \aab)$.   More detail may be
found in Ref. \cite{gll}.

\section{\boldmath Observation of $K_S^0K_L^0$ in $\psi(2S)$ decays and
$J/\psi$ decays}

   There is a longstanding "$\rho\pi$ puzzle" between $J/\psi$ and
$\psi(2S)$ decays in some modes: compared with the corresponding
$J/\psi$ decays, many $\psi(2S)$ decay channels are suppressed
relative to the pQCD predicted "12\% rule"~\cite{BES1}.  Here
$\psi(2S) \rightarrow K_S^0K_L^0$ is observed for the first time in
the BESII 14 million $\psi(2S)$ event sample, and the
branching ratio is used to test the "12\% rule" between $J/\psi$ and
$\psi(2S)$ decays.

    Candidate events are required to have two charged tracks with net
charge zero. The two tracks are assumed to be $\pi^+$ and $\pi^-$, and
the fitted intersection of the two tracks is taken as the $K_S^0$ vertex.
After requiring the $\pi^+\pi^-$ invariant mass within twice the mass
resolution around the nominal $K_S^0$ mass and a $K_S^0$ decay length
in the transverse plane longer than 1 cm, the $K_S^0$ momentum
distribution, shown in Figure ~\ref{fig:psi2s_Ks}, is
obtained. Also shown are the $K_S^0$ mass side band
events and Monte Carlo simulated backgrounds. The signal peak at 1.77
GeV is
fitted with a Gaussian, and the background below the peak is fitted
with an exponential function.  The preliminary branching
ratio obtained is $(5.24 \pm 0.47 \pm 0.48) \times 10^{-5}$. This branching
ratio, together with results for $\psi(2S)\rt \pi^+ \pi^-$ and
$\psi(2S)\rt K^+ K^-$, have been used to extract the relative phase between the
three-gluon and the one-photon annihilation amplitudes of $\psi(2S)$
decays to pseudoscalar meson pairs. It is found that a phase around
$\pm 90^{\circ}$ can explain the result ~\cite{Yuancz}.
\begin{figure}
\resizebox{0.41\textwidth}{!}{%
  \includegraphics{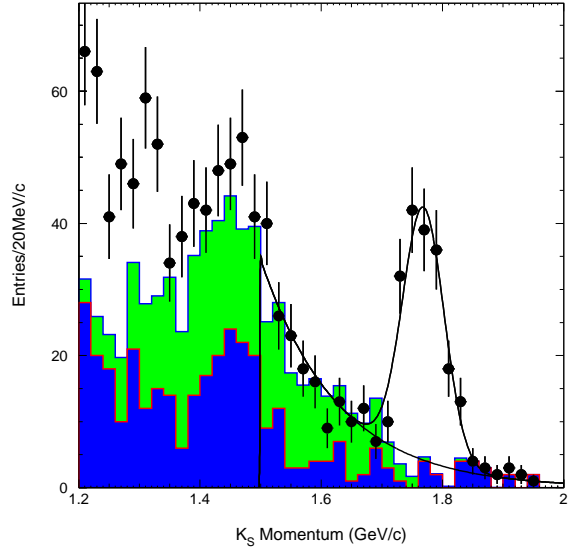}}
\caption{The $K_S^0$ momentum distribution in the $\psi(2S) \rightarrow 
K_S^0K_L^0$. The dots with error bars
are data, the dark shaded histogram is from $K_S^0$ mass side band events, 
and the light shaded histogram is the Monte Carlo simulated background.
The curve shown in the plot is the best fit of the distribution.}
\label{fig:psi2s_Ks}
\end{figure}

\begin{figure}
\resizebox{0.41\textwidth}{!}{%
  \includegraphics{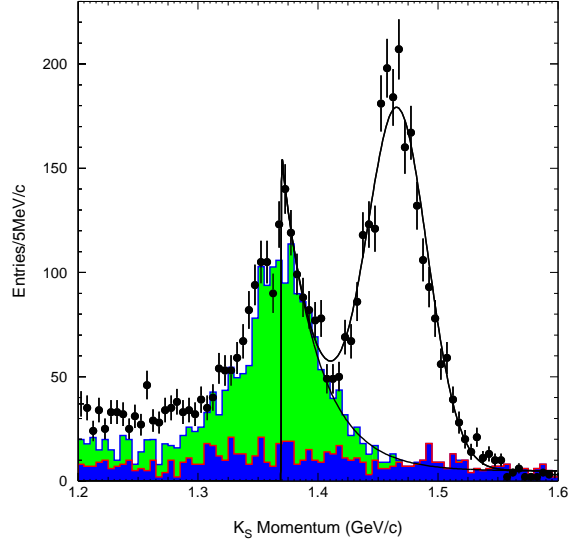}}
\caption{The $K_S^0$ momentum distribution in $J/\psi \rightarrow 
K_S^0K_L^0$.}
\label{fig:jpsi_Ks}
\end{figure}

     With a similar analysis using the BESII 58 million $J/\psi$ events, we also measured the branching ratio of $J/\psi \rightarrow
K_S^0K_L^0$.  The $K_S^0$ momentum distribution in $J/\psi
\rightarrow K_S^0K_L^0$ decays is shown in Figure
~\ref{fig:jpsi_Ks}. The preliminary branching ratio of $J/\psi
\rightarrow K_S^0K_L^0$ is $(1.82\pm 0.04\pm 0.13)\times
10^{-4}$. This result is significantly larger than the world average
of $((1.08\pm 0.14)\times 10^{-4})$ ~\cite{PDG}. Using the branching
ratio of $\psi(2S) \rightarrow K_S^0K_L^0$, and considering the common
errors which cancel out in the calculation of the ratio of the two
branching ratios, one obtains
\begin{equation}
 Q_h=\frac{B(\psi(2S)\rightarrow K_s^0 K_L^0)}{B(J/\psi\rightarrow 
K_s^0K_L^0)}=(28.2\pm3.7)\%.
\label{eq:Qh}
\end{equation}
This number deviates from the pQCD predicted "12\% rule" by more than
four sigma. Most interesting is that this channel is
enhanced in $\psi(2S)$ decays, while in almost all other channels
which deviate from the "12\% rule", $\psi(2S)$ decays are suppressed.
These results are preliminary.  More detail may be found in
Refs.~\cite{psipkskl} and \cite{jpsikskl}.

\section{\boldmath Search for $\psi(3770)$ non-$D\bar{D}$ decay to 
$\pi^+\pi^-J/\psi$}

  The $\psi(3770)$ resonance is believed to be a mixture of $2^3S_1$
and $1^3D_1$ states of the $c\bar{c}$ system ~\cite{Rapidis}. Since
its mass is above open charm-pair threshold and its width is two
orders of the magnitude larger than that of the $\psi(2S)$, it is
thought to decay almost entirely to pure $D\bar{D}$
~\cite{Bacino}. However, recently some theoretical calculations point
out that $\psi(3770)$ could decay to non-$D\bar{D}$ final states
~\cite{Lipkin}.

   Here, we report evidence for $\psi(3770) \rightarrow
\pi^+\pi^-J/\psi$ based on $8.0\pm 0.5$ $pb^{-1}$ of data taken in the
cm energy region around 3.773 GeV with BESII.  Another
source of $\pi^+\pi^-J/\psi$ is from the radiative return process (due
to initial state radiation (ISR)) to the $\psi(2S)$ followed by $\psi(2S)
\rt\pi^+\pi^-J/\psi$.  We developed a new generator {\it isrpsi} which
includes production of $J/\psi$, $\psi(2S)$ and other resonances due
to radiative return. The Monte Carlo simulation includes
leading-log-order radiative return, where the cm energies after ISR
are generated according to Ref. ~\cite{Kuraev}.

To search for the decay of $\psi(3770) \rightarrow \pi^+\pi^-J/\psi$
and $J/\psi \rightarrow e^+e^-$ or $\mu^+\mu^-$,
$e^+e^-\pi^+\pi^-$ and $\mu^+\mu^-\pi^+\pi^-$  candidate events are selected. They are
required to have four charged tracks with zero total charge.

\begin{figure}
\resizebox{0.48\textwidth}{!}{%
  \includegraphics{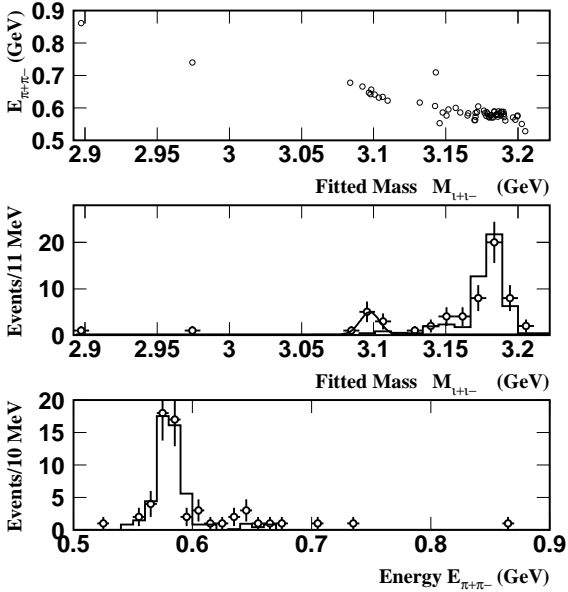}}
\caption{Scatter plot of the measured $\pi^+\pi^-$ 
energies versus the fitted $l^+l^-$ masses. There are two clusters.
The cluster whose energy is around 0.65 GeV is mostly composed of
signal events from $\psi(3770) \rt  \pi^+\pi^-J/\psi$, while the
events whose energies are around 0.57 GeV are due to radiative return
events. The projections are shown in the middle and bottom plots.
The open circles are data, the histograms are the results of 
the Monte Carlo simulation for $\psi(2S) \rightarrow \pi^+\pi^-J/\psi$, 
and the solid smooth curve is the fit to the data.  } 
\label{fig:psipp}
\end{figure}

Figure ~\ref{fig:psipp} shows a scatter plot of the $\pi^+\pi^-$
energies versus the fitted masses of the $l^+l^-$ after a four
constraint fit to the process $\psi(3770) \rightarrow
\pi^+\pi^-J/\psi$.  In the middle sub-figure, the higher mass peak is
due to radiative return to the $\psi(2S)$ followed by $\psi(2S) \rt
\pi \pi J/\psi$.  This peak is shifted to 3.18 GeV because $E_{cm}$ is
set to 3.773GeV in the kinematic fitting.  The fit to this peak yields
a total of $2.2\pm 0.4$ background events near the signal peak at
3.097 GeV, out of the $9.0\pm 3.0$ events.  After background
subtraction, $6.8\pm 3.0$ signal events remain. The branching fraction
for the non-$D\bar{D}$ decay $\psi(3770) \rightarrow \pi^+\pi^-J/\psi$
is measured to be
\begin{equation}
 BF(\psi(3770) \rightarrow \pi^+\pi^-J/\psi) =(0.59\pm0.26\pm0.16)\%,
\label{eq:psipp1}
\end{equation}
where the first error is statistical and the second systematic. Using 
the total width of the $\psi(3770)$ resonance from the PDG  ~\cite{PDG}, this 
branching 
ratio corresponds to a partial width of
\begin{equation}
 \Gamma(\psi(3770) \rightarrow \pi^+\pi^-J/\psi) =(139\pm 61\pm 41)
 {\rm ~keV}.
\label{eq:psipp2}
\end{equation}
More detail may be found in Ref.~\cite{gongr}.  These results are preliminary.

\end{document}